# Large Work Extraction and the Landauer limit in the Continuous Maxwell Demon

M. Ribezzi-Crivellari[a,b] and F. Ritort[a,c]

[a]Condensed Matter Physics Department, University of Barcelona, C/Marti i Franques s/n, 08028 Barcelona, Spain.

[b]Laboratoire de Biochimie, Institute of Chemistry, Biology and Innovation (CBI), UMR 8231, ESPCI Paris/CNRS, PSL Research University, 10 rue Vauquelin, 75231-Paris Cedex 05, France

[c]CIBER-BBN de Bioingeniería, Biomateriales y Nanomedicina, Instituto de Sanidad Carlos III, Madrid, Spain.

**The relation between entropy and information dates back to the classical Maxwell demon (MD) paradox [1], a thought experiment proposed in 1867 by J. C. Maxwell to violate the second law of thermodynamics. A variant of the classical MD is the Szilard engine proposed by L. Szilard in 1926 in which the demon observes, at a given time, the compartment occupied by a single molecule in a vessel and extracts work by operating a pulley device. Here we introduce the Continuous Maxwell Demon (CMD), a device capable of extracting arbitrarily large amounts of work per cycle by repeated measurements of the state of a system, and experimentally test it in single DNA hairpin pulling experiments. In the CMD the demon monitors the state of the DNA hairpin (folded or unfolded) by observing it at equally spaced time intervals but extracts work only when the molecule changes state. We demonstrate that the average maximum work per cycle that can be extracted by the CMD is limited by the information-content of the stored sequences, in agreement with the second law. Work extraction efficiency is found to be maximal in the large information-content limit where work extraction is fueled by rare events.**

In the Szilard engine the demon performs a one-bit measurement by observing, at a given time, the compartment $(V_0, V_1)$ occupied by a single molecule in a vessel of volume $V$ at temperature $T$ (Figure 1A). The engine operates as follows: if the molecule occupies the left compartment ($V_0$), a pulley device extracts the mean work $W_0 = -k_B T \log P_0$ where $P_0 = V_0/V$ is the probability of the molecule observed in the left compartment; if it occupies the right compartment ($V_1 = V - V_0$) the mean extracted work equals $W_1 = -k_B T \log P_1$ with $P_1 = 1 - P_0$. The average work per cycle that can be extracted in the classical MD equals

$$W_{max}^{MD} = P_0 W_0 + P_1 W_1 = -k_B T (P_0 \log P_0 + P_1 \log P_1) \quad . \tag{1}$$

It is maximal for $P_0 = P_1 = 1/2$, $W_{max}^{MD} \leq W_L = k_B T \log 2$, $W_L$ being the Landauer limit. The resolution of the MD paradox, i.e. the fact that the engine can fully convert heat into work without any other change, came from the thermodynamics of data processing. Half a century ago it was shown that any irreversible logical operation, such as bit erasure, requires energy consumption typically on the order of $k_B T$ [2,3]. In general, $W_{max}^{MD}$ equals the information-content $I$ of one bit, $I = -P_0 \log P_0 - P_1 \log P_1$, restoring the second-law inequality,

$$W \leq W_{max}^{MD} = k_B T I \qquad (2)$$

with $W$ the mean extracted work. Subsequent developments in experimental physics, often in combination with the theory of fluctuation theorems and information feedback [4-8] have provided experimental realizations and models of the MD that have tested Eq.2 and the Landauer limit [9-16].

Here we introduce the continuous MD (CMD), a conceptually new information-to-energy conversion device that takes advantage of extracting work from rare events. The CMD is exemplified in Figure 1B. The demon monitors the motion of the molecule by observing it at equally spaced time intervals $\tau$ but extracts work only when the molecule changes compartment. A work-extraction cycle starts with an initial observation of the compartment occupied by the molecule that can be 0 (left compartment) or 1 (right compartment), followed by a series of repeated measurements every $\tau$ until the molecule changes compartment. We classify cycles into two classes, 0-cycles and 1-cycles, for cycles starting with initial outcome measurement equal to 0 and 1, respectively. This is then followed by a series of $n - 1$ repeated measurements of equal outcome made at times $\tau, 2\tau, 3\tau, \ldots, (n-1)\tau$ until the molecule changes compartment at time $n\tau$: $(0 \to 1)$ for a 0-cycle and $(1 \to 0)$ for a 1-cycle. For cycles containing multiple observations the demon stores in memory the $(n+1)$-bits sequence containing $n$-equal measurement outcomes plus the final (different) outcome. Stored sequences are defined as $\mathbf{0}_n = \{\overbrace{0,\ldots 0}^{n}, 1\}$ for 0-cycles and $\mathbf{1}_n = \{\overbrace{1,\ldots 1}^{n}, 0\}$ for 1-cycles ($n \geq 1$). The mean work per cycle depends on the last bit in the cycle and equals $W_1 = -k_B T \log P_1$ for 0-cycles and $W_0 = -k_B T \log P_0$ for 1-cycles. The average maximum work per cycle that can be extracted by the CMD is independent of the time interval $\tau$ and is given by,

$$W_{max}^{CMD} = P_0 W_1 + P_1 W_0 = -k_B T (P_0 \log P_1 + P_1 \log P_0) \quad , \qquad (3)$$

with $P_0, P_1$ the probabilities of 0-cycles and 1-cycles (i.e. determined by the initial bit in the cycle). Albeit similar to Eq.(2) the functional dependence of $W_{max}^{CMD}$ on $P_0$ in Eq.(3) is different (Figure 1C). In particular $W_{max}^{CMD}$ in Eq.(3) is minimal (rather than maximal) for $P_0 = P_1 = 1/2$, the Landauer limit being a lower (rather than an upper) bound, $W_{max}^{CMD} > W_L = k_B T \log 2$. Moreover, $W_{max}^{CMD}$ diverges in the limits $P_0 \to 0,1$ showing that the CMD can extract arbitrarily large amounts of work. Should this violate the second law inequality Eq.(2)? The answer is negative because when $W_{max}^{CMD}$ diverges it also does the information-content $I$ of the stored sequences $\mathbf{0}_n, \mathbf{1}_n$. In fact, work extraction cycles in the CMD require storing multiple bit sequences of the class $\mathbf{0}_n, \mathbf{1}_n$ (Figure 1B), which length diverges in the limits $P_0 \to 0,1$ when the molecule changes compartment after many measurements. This is in contrast with the classical MD where work extraction cycles just require a single bit measurement (Figure 1A). For a two-state system, with relaxation rate equal to $R$, the average information-content $I$ in the stored multiple-bit sequences can be exactly computed (Sec. S1 in Supp. Info). It is given by $I(\tau) = I_{min} + I_1(\tau)$, where $I_{min} > 0$ is the minimum information-content and $I_1(\tau) > 0$ is a monotonically decreasing function of $\tau$ such that $I_1(\tau \to \infty) = 0$. The expressions for $I_{min}, I_1(\tau)$ are given by,

$$I_{min} = -\frac{P_0}{P_1}\log(P_0) - \frac{P_1}{P_0}\log(P_1) - P_0 \log P_1 - P_1 \log P_0 \qquad (4)$$

$$I_1(\tau) = -\frac{P_0(P_0+P_1 e^{-R\tau})}{P_1(1-e^{-R\tau})}\log\left(1+\frac{P_1}{P_0}e^{-R\tau}\right) - \frac{P_1(P_1+P_0 e^{-R\tau})}{P_0(1-e^{-R\tau})}\log\left(1+\frac{P_0}{P_1}e^{-R\tau}\right) - \log(1-e^{-R\tau}) - \left(\frac{P_0 \log(P_0)}{P_1} + \frac{P_1 \log(P_1)}{P_0}\right)\frac{e^{-R\tau}}{1-e^{-R\tau}} \qquad (5)$$

The minimum information-content $I_{min}$ is obtained in the limit $R\tau \gg 1$, where $I_1(\tau) = -e^{-R\tau}(P_0 \log(P_0)/P_1 + P_1 \log(P_1)/P_0) + O(e^{-2R\tau})$; whereas a diverging value is obtained for $R\tau \ll 1$ : $I_1(\tau \to 0) = -\log(R\tau) + 1 + \frac{P_0^2 \log(P_0)}{P_1} + \frac{P_1^2 \log(P_1)}{P_0}$ giving, $I_{max}(\tau) = -\log(R\tau) + 1 - \log P_0 - \log P_1$. The chain of inequalities follows,

$$W \le W_{max}^{CMD} < k_B T I_{min} < k_B T I(\tau) < k_B T I_{max}(\tau) \quad . \qquad (6)$$

We stress that $W_{max}^{CMD}$ is independent of $\tau$. The lowest value of $I_{min}$ in Eq.(4,6) is obtained for $P_0 = P_1 = 1/2$, $I_{min} = 3\log(2)$, with stored sequences containing 3 bits in average. This in contrast to the classical MD where the information-content of one-bit sequences equals $\log(2)$. In fact, the CMD must store, at least, two bits per cycle (the first bit defining the cycle class, the last bit closing the cycle when the molecule changes compartment). The efficiency of the CMD is defined by the ratio between the maximum extracted work $W_{max}^{CMD}$ (c.f. Eq.(3)) and the energy required to erase the stored sequences, $Q = k_B T I$. Maximum efficiency $\epsilon_{max}$ is obtained in the limit $\tau \to \infty$:

$$\epsilon_{max} = \frac{W_{max}^{CMD}}{k_B T I_{min}} = \left(1 + \frac{P_0 \log(P_0)/P_1 + P_1 \log(P_1)/P_0}{P_0 \log(P_1) + P_1 \log(P_0)}\right)^{-1} < 1 \quad . \qquad (7)$$

Interestingly, for $P_0 = P_1 = 1/2$ the efficiency is minimal in the CMD ($\epsilon_{max} = 1/2$), whereas it is maximal in the classical MD ($\epsilon_{max} = 1$). Instead, in the limits $P_0 \to 0,1$ the CMD yields maximum efficiency $\epsilon_{max} \to 1$ . The behavior of $W_{max}^{MD}, W_{max}^{CMD}, I(\tau), I_{min}, \epsilon_{max}$ is shown in Figure 1C (continuous lines). Note that for uncompressed sequences (i.e. that contain redundant information) the efficiency is lower than Eq.7 (Sec. S2 in Supp. Info.).

Equation (5) shows that information-content diverges in the continuum-time limit $\tau \to 0$ when sequences contain an arbitrary large number of bits. However it does logarithmically, $I_{max}(\tau) \to -\log(R\tau)$, rather than linearly with the number of bits, $I_{max}(\tau) \to 1/R\tau$, showing that stored sequences are highly redundant for $\tau \ll 1/R$. A similar problem is found in data compression where information can be encoded using fewer bits than in the original representation [17]. In general, the information-content of sequences, storing the outcome of measurements repeated at time intervals $\tau$, diverges logarithmically for $\tau$ smaller than the decorrelation time.

Recent technological advancements have made possible the practical implementation of Szilard engines [9,10,13]. Here we report a novel room-temperature, nanoscale Szilard engine composed of a single DNA molecule manipulated by optical tweezers (Figure 2A). In our experiments a single DNA hairpin is tethered between two micron-sized plastic beads (Materials and Methods). The force applied on the molecule is controlled and measured varying the position of the optical trap $\lambda$. Under a suitable force the molecule will exhibit spontaneous fluctuations between the folded and the unfolded states. This is equivalent to the initial state of the Szilard engine, where the molecule can freely transition from the left to the right compartments. In the classical MD the state of the DNA hairpin is observed at a given time and, depending on whether it is folded or unfolded, a pulling protocol to extract work implemented using the DNA as a pulley device (Figure 2B, top). In the CMD the state of the hairpin is monitored every time $\tau$ and the same protocol is implemented when the molecule changes state (Figure 2B, bottom). The work extraction protocol is cyclic i.e. the control parameter $\lambda$ is first driven away from its initial value $\lambda_0$ and then driven back to it. While the first, forward, part of the protocol is swift (irreversible) the second, reverse part, is slow (adiabatic). Less work is consumed in the forward part of the protocol than what can be extracted in the reverse part yielding a net amount of extracted work (Sec. S3 in Supp. Info.). It is possible to show that dissipation losses during the work extraction cycle are negligible (Sec. S4 in Supp. Info.) The spontaneous hopping events observed in the adiabatic part of the protocol (Figure 2B, red traces) are the equivalent of the collisions against the inserted wall in the one-molecule Szilard gas, and contribute to net work extraction. Let the bit 0 (1) denote the folded (unfolded) state. We have $P_0 = \frac{1}{1+e^{\phi}}$ and $P_1 = \frac{1}{1+e^{-\phi}}$ with $\phi$ being the equilibrium free energy difference (in $k_B T$ units) between the folded and the unfolded states at $\lambda_0$ : $\phi = (G_0 - G_1)/k_B T = \Delta G/k_B T = -\log\left(\frac{P_0}{P_1}\right)$.

A comparison of the results obtained for the classical MD and the CMD are shown in Figure 3. In Figure 3A we show hopping traces at different equilibrium conditions ($-1.4 < \phi < 1.4$). Figures 3B,C show the extracted work distributions together with the mean extracted work $W_{max}$ (circles) measured over many cycles in the classical MD and the CMD (work distributions being independent of the value of $\tau$). The difference between both work distributions and the mean extracted work is clear. Distributions are bimodal for $\phi \neq 0$, with larger amounts of work extracted from the CMD as compared with the classical MD. In particular, the Landauer limit (vertical dashed line) is an upper bound for the classical MD but a lower bound for the CMD. The Landauer limit is met in both cases at the coexistence point $P_0 = P_1 = 1/2, \phi = 0$. The measured values for $W_{max}^{MD}, W_{max}^{CMD}$ and $\epsilon_{max}$ are shown in Figure 1C and compared to the theoretical predictions Eq. (1,3,7).

It is remarkable that the Landauer limit is a lower bound of the mean extracted work for the CMD (rather than an upper bound). The large amount of work that can be extracted in the CMD in the limit $|\phi| \gg 1$ comes at the price of the long time required to observe the system leaving the most probable state. In fact, the average cycle time in the CMD, $t_c^{CMD}$, is given by (Sec. S1 in Supp. Info),

$$t_c^{CMD}/\tau = \frac{1}{1-e^{-R\tau}}\left(\frac{1+e^{2\phi}}{e^{\phi}}\right) + 1 > 3 \quad . \tag{8}$$

From Eqs. (1,3) we obtain for the relative power between the two cases, $P_{CMD}/P_{MD} \to 1 - O(1/|\phi|)$ in the limit $|\phi| \gg 1$, showing that the extracted power of both engines is asymptotically the same. The advantage of the CMD with respect to the classical case is the large amount of work per cycle that can be extracted in the CMD (Fig. 3C). This originates from the potentially unlimited information-content of the stored sequences, which in the one-bit classical MD cannot exceed $\log 2$. Moreover the CMD captures rare dynamical events that deliver large amounts of work, whereas the classical MD captures typical events that only yield a moderate amount (Sec. S5 in Supp. Info.). We note that the fact that the average power is asymptotically the same does not imply that the power distributions for finite time are equal. In fact the work distributions obtained for the classical MD and the CMD (Figs. 3B and 3C) are already different meaning that the distributions of power for finite time will be different too. The analytical calculation of such power distributions requires applying large deviation theory as has been done for the case of fluctuating efficiencies in heat engines [18].

Our calculation of the information content, based on the Shannon entropy of the stored sequences, follows the spirit of the path thermodynamics approach applied to other nonequilibrium systems [19]. Although other information theoretical quantities introduced in the field of dynamical systems might be suitable to analyze our experimental data [20,21], all probably give the same result in the rare events limit $|\phi| \gg 1$. Furthermore our information-content calculation might be extended to other nonequilibrium situations where work extraction is fuelled by rare dynamical events, e.g. in the context of fluctuation theorems for repeated feedback [22].

Finally, the CMD might be applicable to other contexts, such as biological and quantum systems [23,24]. For example, the CMD might be relevant in regulatory biological networks when the concentration of molecular species reach a given threshold such as during the generation and transmission of action potential signals across the cell membrane and signal transduction processes in general [25]. In all these cases there is a continuously monitored physical variable and the cell responds when such variable abruptly changes. For example, the action potential signal is generated when the potential difference across the cell membrane reaches a threshold. In this case the work extracted would be equal to the free energy required to produce that action potential signal at a specific membrane location, whereas information would be contained in the stored trains of bits defined by an inactive (0) or active (1) membrane potential relative to the threshold. One cannot avoid thinking whether the information-to-energy conversion in the CMD might have consequences in self-organization and selection processes taking place during the evolution of biological matter [26]. The astonishing complexity of living matter might be seen as the outcome, over long evolutionary timescales, of a large work extraction process (used to build new and emergent high free energy structures) in environments suitable to store large amounts of information utterly erased by noise and randomness.

**Acknowledgements.** We acknowledge financial support from Grants 308850 INFERNOS, 267862 MAGREPS (FP7 EU program) FIS2013-47796-P, FIS2016-80458-P (Spanish Research Council) and Icrea Academia prize 2013 (Catalan Government).

**Materials and methods**

Experiments were performed with a highly stable miniaturized optical tweezers setup [27]. Two tightly focused counter-propagating laser beams (P = 200 mW, $\lambda$ =845 nm) are used to create a single optical trap. Experiments are performed in a microfluidics chamber. Force measurements, based on detection of light momentum changes, are performed using Position Sensitive Detectors (PSD) to measure the deflection of the laser beam after it interacts with the trapped object. The position of the trapping beam defines the control parameter $\lambda$ that is measured by diverting ~ 8% of each laser beam to a secondary PSD. The instrument has a resolution of 0.1 pN and 1 nm at a 1 kHz acquisition rate. In the experiments a molecular construct consisting of a 20bp DNA hairpin flanked by two short 29bp DNA handles (labelled with biotin and digoxigenin) was tethered between two polystyrene beads. One type of bead is optically trapped whereas the other bead is immobilized on the tip of a micropipette by air suction. Each handle can selectively bind to either streptavidin (1.87 μm, Spherotech) or anti-digoxigenin coated beads (3.0–3.4 μm Kisker Biotech). The synthesis protocols for short (20 bp) DNA hairpins have been previously described [28]. All experiments were performed at 25°C in a buffer containing: 10 mM Tris, 1 mM EDTA, 1 M NaCl, 0.01% NaN3 (pH 7.5).

**Figure Captions**

**Figure 1. Classical MD vs CMD.** (A) In the classical MD a single observation is made and depending on the measurement outcome (0,1) a work extraction process implemented which yields mean work values $(W_0, W_1)$, respectively. (B) In the CMD multiple observations are made every $\tau$ and a work extraction process implemented when the molecule changes compartment $(1 \rightarrow 0, 0 \rightarrow 1)$ yielding the mean work values $(W_0, W_1)$, respectively. (C) Work, information-content and efficiency as a function of $P_0$ for the classical MD and the CMD. Mean work (classical MD: pink up-triangles; CMD: blue diamonds) obtained from the experiments (see Figures 2,3) and theoretical prediction from Eqs.(1,3) (pink and blue lines). The Landauer limit $W_L = k_B T \log 2$ is also shown (horizontal line). Information-content in the CMD for different values of $R\tau$ (Eqs(4,5,6)) from $R\tau = 0.1$ to $R\tau = \infty$ (light to dark green lines). Efficiency in the CMD (orange diamonds) from Eq.(7) (scale in the right axis).

**Figure 2. Single-molecule tests for the classical MD and CMD.** (A) Schematics of the steps in a work extraction cycle in a DNA hairpin pulling experiment. If the DNA hairpin is found in the folded state (step 1) the trap is suddenly approached by a distance $\Delta\lambda$ and the force relaxed (step 2). A work extraction process is implemented where the trap slowly recovers the original position $\lambda$ (steps 3-4). Work is extracted every time the molecule spontaneously unfolds. An equivalent cycle is implemented if the molecule is found in the unfolded state. (B) Implementation of the work extraction process in the classical MD (upper traces) and the CMD (lower traces). Typical force-time traces previous to the work-extraction process (hopping trace, blue) and during the work-extraction adiabatic process (tilted hopping trace, red). The characteristic trap-position vs. time stepwise function is shown in black.

**Figure 3. Extracted work distributions.** (A) Hopping traces at different values of $P_0$. Extracted work histograms in the classical MD (panel B) and in the CMD (panel C) at the corresponding values of $P_0$. Histograms are fit to double Gaussian distributions (continuous lines in color). The black circles are the mean of the extracted work distributions and the black lines are the theoretical prediction, Eq.(1) for the classical MD (panel B) and Eq.(3) for the CMD (panel C).

# Figure 1

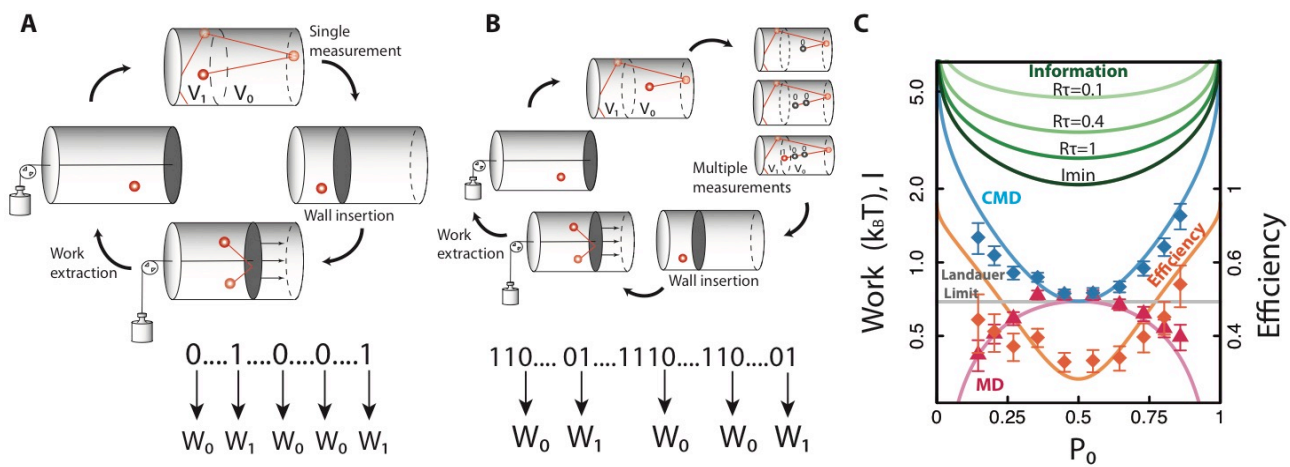

# Figure 2

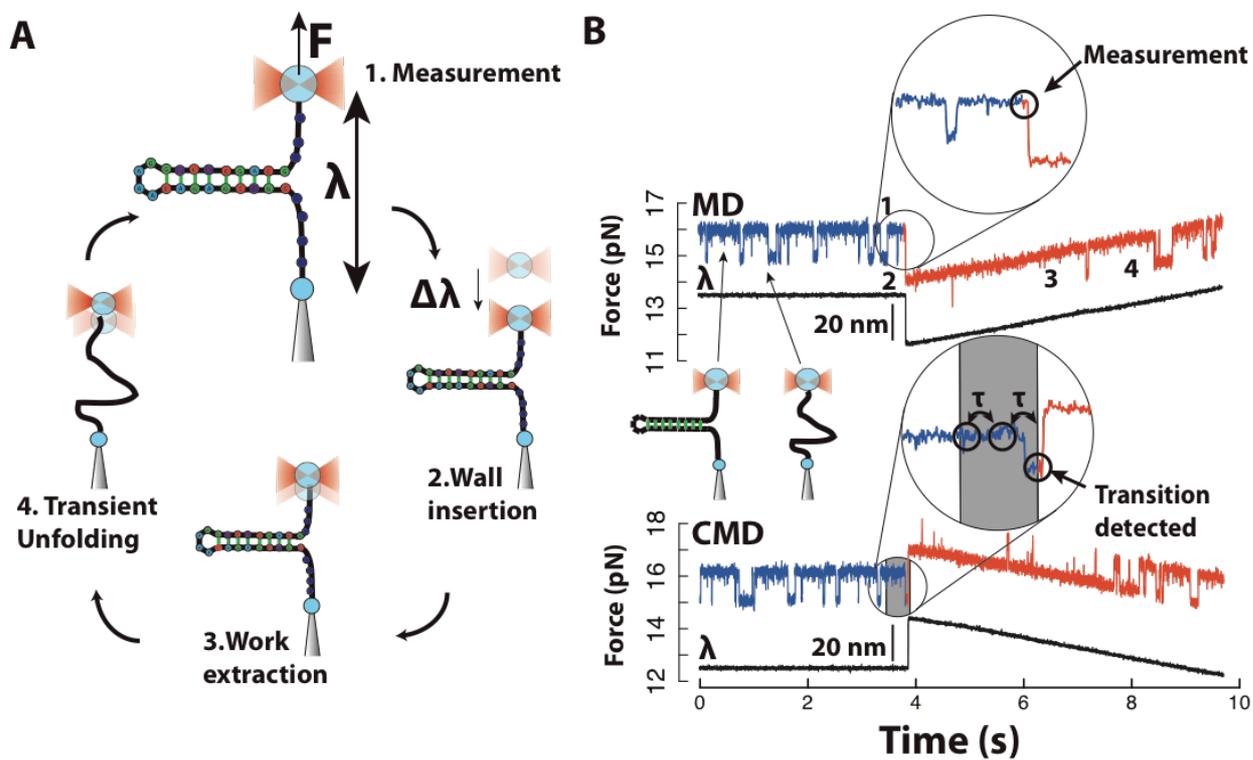

# Figure 3

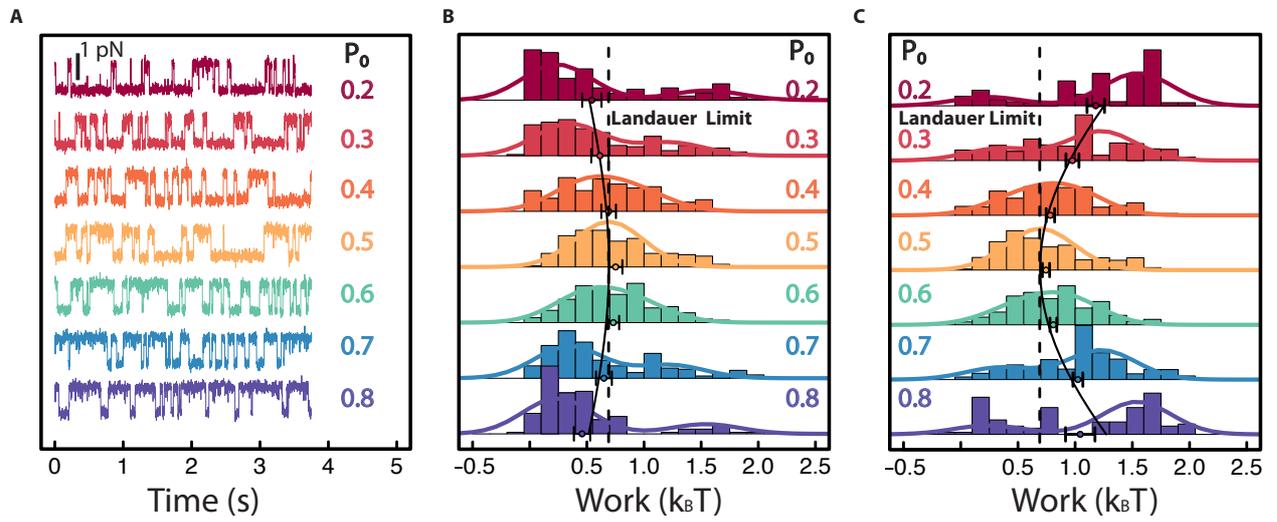

# Supplementary information of "Continuous Maxwell Demon and the Landauer limit"

M. Ribezzi-Crivellari[a,b] and F. Ritort[a,c]

[a]Condensed Matter Physics Department, University of Barcelona, C/Marti i Franques s/n, 08028 Barcelona, Spain.

[b]Laboratoire de Biochimie, Institute of Chemistry, Biology and Innovation (CBI), UMR 8231, ESPCI Paris/CNRS, PSL Research University, 10 rue Vauquelin, 75231-Paris Cedex 05, France

[c]CIBER-BBN de Bioingeniería, Biomateriales y Nanomedicin a, Instituto de Sanidad Carlos III, Madrid, Spain.

**S1: Information-content calculation for the two-states model**

The two-compartment single-particle system and the single DNA pulling experiment can be modeled by a two-states system $\sigma = 0,1$ with kinetic rates $k_{\sigma \to \sigma'}$ for the system to change state $\sigma \to \sigma'$. The equilibrium occupancies are given, $P_0 = k_{1\to 0}/R$ ; $P_1 = k_{0\to 1}/R$, where $R = k_{1\to 0} + k_{0\to 1}$ is the full relaxation rate. Rates satisfy the detailed balance condition $k_{0\to 1}/k_{1\to 0} = P_1/P_0$ with $P_0 = \frac{1}{1+e^\phi}$ ; $P_1 = \frac{1}{1+e^{-\phi}}$. The dimensionless quantity $\phi$ stands for the free energy difference (in $k_B T$ units) between states 0 and 1,

$$\phi = (G_0 - G_1)/k_B T = \Delta G/k_B T = -\log\left(\frac{P_0}{P_1}\right) \quad . \tag{S1.1}$$

For the two-compartment model $\phi = \log(V_1/V_0)$ whereas in the single DNA pulling experiment $\phi$ equals the free energy difference between the folded and unfolded states. Let $W_t(\sigma|\sigma')$ be the conditional probability of the system being in state $\sigma$ at time $t$ if it is in state $\sigma'$ at time 0. It satisfies the following equation,

$$\frac{\partial W_t(\sigma|\sigma')}{\partial t} = \sum_{\sigma''}[\,k_{\sigma''\to\sigma}W_t(\sigma''|\sigma') - k_{\sigma\to\sigma''}W_t(\sigma|\sigma')] \tag{S1.2}$$

where $\sum_\sigma W_t(\sigma|\sigma') = 1$. The equations are readily solved

$$W_t(1|0) = \frac{k_{0\to 1}}{R}(1 - e^{-Rt}) \; ; \; W_t(0|0) = 1 - W_t(1|0) \tag{S1.3}$$

$$W_t(0|1) = \frac{k_{1\to 0}}{R}(1 - e^{-Rt}) \; ; \; W_t(1|1) = 1 - W_t(0|1) \tag{S1.4}$$

The calculation of the information-content $I$ of the stored sequences follows basic steps in information theory. The stored sequences are defined by $\mathbf{0}_n = \{\overbrace{0,\ldots 0}^{n}, 1\}$ for 0-cycles and $\mathbf{1}_n = \{\overbrace{1,\ldots 1}^{n}, 0\}$ for 1-cycles ($n \geq 1$). The first bit in the cycle specifies the first measurement outcome and the corresponding cycle class, whereas the last bit indicates that the system has changed state and the measurement outcome changes bit (1 for 0-cycles and 0 for 1-cycles). Measurements are made every $\tau$ and we assume measurements are free of error. The probability of the sequences $\mathbf{0}_n, \mathbf{1}_n$ are defined as $p_n, q_n$ respectively. These are given by, $p_n = P_0 W_\tau^{n-1}(0|0) W_\tau(1|0)$ and $q_n = P_1 W_\tau^{n-1}(1|1) W_\tau(0|1)$, and satisfy the condition, $\sum_{n=1}^{\infty}(p_n + q_n) = 1$. The information-content of the stored sequences is given by

$$I = -\sum_{n=1}^{\infty}(p_n \log p_n + q_n \log q_n) \quad . \tag{S1.5}$$

Simple algebra shows that $I$ it can be decomposed into two terms, $I = I(C) + I(S|C)$ where $C$ stands for the cycle class and $S$ stands for the stored sequence. The term $I(C)$ equals the information content related to the cycle class, i.e. the information-content of the first bit in the stored sequences $S$. Therefore $I(C) = -P_0 \log P_0 - P_1 \log P_1$ which equals the information content in the classical MD. The term $I(S|C)$ is the information-content in the ensemble of sequences $S$ conditional to the cycle class or the first bit measurement. It is given by,

$$I(S|C) = -\left\{P_0 \left(\frac{W_\tau(0|0)}{W_\tau(1|0)} \log W_\tau(0|0) + \log W_\tau(1|0)\right) + P_1 \left(\frac{W_\tau(1|1)}{W_\tau(0|1)} \log W_\tau(1|1) + \log W_\tau(0|1)\right)\right\} \quad . \tag{S1.6}$$

Substituting Eqs.(S1.3,S1.4) in the previous expressions gives the result reported in the main text, $I = I(C) + I(S|C) = I_{min} + I_1(\tau)$ with $I_{min}, I_1(\tau)$ given in Eq. (4,5),

$$I_{min} = -\frac{P_0}{P_1}\log(P_0) - \frac{P_1}{P_0}\log(P_1) - P_0 \log P_1 - P_1 \log P_0 \tag{S1.7}$$

$$I_1(\tau) = -\frac{P_0(P_0 + P_1 e^{-R\tau})}{P_1(1-e^{-R\tau})}\log\left(1 + \frac{P_1}{P_0}e^{-R\tau}\right) - \frac{P_1(P_1 + P_0 e^{-R\tau})}{P_0(1-e^{-R\tau})}\log\left(1 + \frac{P_0}{P_1}e^{-R\tau}\right) - \log(1 - e^{-R\tau}) - \left(\frac{P_0 \log(P_0)}{P_1} + \frac{P_1 \log(P_1)}{P_0}\right)\frac{e^{-R\tau}}{1-e^{-R\tau}} \quad . \tag{S1.8}$$

The $\tau$ dependent contribution $I_1(\tau)$ diverges logarithmically in the limit $\tau \to 0$,

$$I_1(\tau \to 0) = -\log(R\tau) + 1 + \frac{P_0^2 \log(P_0)}{P_1} + \frac{P_1^2 \log(P_1)}{P_0} + O(R\tau) \quad . \tag{S1.9}$$

In a similar way one can calculate the average cycle time in the CMD, $t_c^{CMD}$. This is equal to the average number of steps in one state before the system changes state multiplied by the time interval $\tau$. In other words $t_c^{CMD}/\tau$ is equal to the average length of the stored sequences, $\mathbf{0}_n, \mathbf{1}_n$. It is given by,

$$t_c^{CMD}/\tau = \sum_{n=1}^{\infty}(n+1)(p_n + q_n) = 1 + \sum_{n=1}^{\infty}n(p_n + q_n) =$$
$$= 1 + \frac{P_0}{W_\tau(1|0)} + \frac{P_1}{W_\tau(0|1)} = \frac{1}{1-e^{-R\tau}}\left(\frac{1+e^{2\phi}}{e^\phi}\right) + 1 \quad , \tag{S1.10}$$

which for finite $R, \tau$ is greater than 3 and is given by Eq.(8) in the main text. For the relative power between the CMD and the classical MD we first calculate the average power in the CMD. To optimize the extracted power in the CMD it is natural to take the limit $\tau \to 0$ in Eq.(S1.10). Because $W_{max}^{CMD}$ is independent of $\tau$, maximum power is obtained in the limit $\tau \to 0$. Equation (S1.10) gives the leading behavior,

$$t_c^{CMD} \to \frac{1}{R}\left(\frac{1+e^{2\phi}}{e^\phi}\right) \qquad . \qquad (S1.11)$$

The average power is then given by,

$$P_{CMD} = \frac{W_{max}^{CMD}}{t_c^{CMD}} = k_B T R \frac{\left(\frac{\log(1+e^{-\phi})}{1+e^\phi} + \frac{\log(1+e^\phi)}{1+e^{-\phi}}\right)}{\left(\frac{1+e^{2\phi}}{e^\phi}\right)} \qquad . \qquad (S1.12)$$

In the classical MD work extraction can be applied repeatedly, let us say every relaxation time $1/R$, yielding the optimal power,

$$P_{MD} = \frac{W_{max}^{MD}}{1/R} = k_B T R \left(\frac{\log(1+e^\phi)}{1+e^\phi} + \frac{\log(1+e^{-\phi})}{1+e^{-\phi}}\right) \qquad (S1.13)$$

In the limit $|\phi| \gg 1$ we have, $P_{CMD} = k_B T R |\phi| e^{-|\phi|}\left(1 + O(e^{-|\phi|})\right)$ and $P_{MD} = k_B T R |\phi| e^{-|\phi|}\left(1 + O(1/|\phi|)\right)$ giving the result in the text: $P_{CMD}/P_{MD} \to 1 - O(1/|\phi|)$.

**S3 : Mean work per cycle in a Szilard engine, the adiabatic case**

In order to understand work extraction from the Szilard Engine we shall compute the work extracted per cycle in a limit case i.e. when the forward (FW) part of the protocol is instantaneous while the reverse part of the protocol (RV) is adiabatic. Likewise in the main text, 0 (1) denotes de folded (unfolded) state respectively. The initial value of the control parameter, $\lambda_0$, will determine the probability of finding the molecule in the folded state, $P_0$, or in the unfolded state, $P_1$. The logarithm of the ratio of these two probabilities is denoted by $\phi$ and equals the free energy difference between the folded and the unfolded state,

$$\phi = -\log\left(\frac{P_0}{P_1}\right) = -\log\left(\frac{e^{-\beta G_0}}{e^{-\beta G_1}}\right) = \beta(G_0 - G_1) = \beta \Delta G \quad . \qquad (S3.1)$$

Let us assume the molecule is initially in the unfolded state 1. The control parameter $\lambda$ will be raised instantaneousely to the value $\lambda_H > \lambda_0$. During this transition the system will be confined to the unfolded free energy branch $G_1$ and the total work performed *on* the system will be:

$$W_1^{FW} = G_1(\lambda_H) - G_1(\lambda_0) \qquad (S3.2)$$

After reaching $\lambda_H$ the protocol will be reversed and the control parameter will be returned adiabatically to $\lambda_0$. Along this part of the protocol the work performed *on* the system will be :

$$W_1^{RV} = G(\lambda_0) - G(\lambda_1) \quad , \tag{S3.3}$$

where the full equilibrium free energy $G(\lambda)$ and the partial free energies $G_1(\lambda), G_0(\lambda)$ are related by :

$$G(\lambda) = -k_B T \log\left(e^{-\beta G_0(\lambda)} + e^{-\beta G_1(\lambda)}\right) \quad . \tag{S3.4}$$

Now if $\lambda_H \gg \lambda_0$ thermodynamic stability implies $G_0(\lambda_H) \gg G_1(\lambda_H)$ and as a consequence we can approximate:

$$G(\lambda_H) = -k_B T \log\left(e^{-\beta G_0(\lambda_H)} + e^{-\beta G_1(\lambda_H)}\right) \approx G_1(\lambda_H) \tag{S3.5}$$

and

$$W_1^{RV} \approx G(\lambda_0) - G_1(\lambda_H) \quad . \tag{S3.6}$$

This allows us to compute the maximum average work per cycle as a function of $\phi$ :

$$W_1 = W_1^{RV} + W_1^{FW} \approx$$

$$\approx G(\lambda_0) - G_1(\lambda_0) = -k_B T \log\left(e^{-\beta G_0(\lambda_0)} + e^{-\beta G_1(\lambda_0)}\right) - G_U(\lambda_0) =$$

$$= -k_B T \log\left(e^{-\beta G_1(\lambda_0)}(1 + e^{-\beta \Delta G(\lambda_0)})\right) - G_1(\lambda_0) = -k_B T \log(1 + e^{-\phi}) \quad . \tag{S3.7}$$

A similar computation can be performed if the molecule is initially found in the folded state, replacing $\lambda_H$ with $\lambda_L < \lambda_0$. In this case the maximum average work per cycle will be given by:

$$W_0 \approx -k_B T \log(1 + e^{\phi}) \tag{S3.8}$$

Equations (S3.7) and (S3.8) are the equivalent of the two-compartment ideal gas model given in the text with the transformation $P_0 = V_0/V = 1/(1 + e^{\phi})$. Summing up, the maximum average work per cycle in the classical MD will be given by,

$$W_{max}^{MD} = P_0 W_0 + P_1 W_1 = k_B T \left(\frac{\log(1+e^{\phi})}{1+e^{\phi}} + \frac{\log(1+e^{-\phi})}{1+e^{-\phi}}\right) \tag{S3.9}$$

whereas for the CMD,

$$W_{max}^{CMD} = P_0 W_1 + P_1 W_0 = k_B T \left(\frac{\log(1+e^{-\phi})}{1+e^{\phi}} + \frac{\log(1+e^{\phi})}{1+e^{-\phi}}\right) \tag{S3.10}$$

which are Eq.(1,3) in the main text.

**S4 : Dissipation losses during the work extraction cycle.**

In our experimental protocol, the 'fast change' is performed at a pulling speed of 2 nm/ms and takes approximately 10 ms. On such timescales the molecule remains in a specific folding state during the transition. As a consequence the main source of dissipation is due to the friction affecting the plastic bead used in the manipulation, an effect that is intrinsically captured in our measurements. In Figure S4 we demonstrate that this effect is negligible. Figure S4A shows a typical realization of our pulling protocol as a function of time during a work extraction cycle. The force is directly measured by detecting changes in the linear momentum of the trapping beam. Fig S4B shows the same data in terms of the control parameter λ. Here the orange points correspond to the first and fast part of the pulling (i.e. when the position of the trap is rapidly changed) whereas the red line corresponds to the second and slow part of the pulling cycle (i.e. when the original position of the trap is adiabatically recovered during the work extraction cycle). Finally the green data correspond to the partial equilibrium in the folded state as estimated from the second slow part of the pulling work extracting cycle.

The partial equilibrium corresponds to the state of the system when configurations are constrained to the folded basin but the mechanical and thermodynamic variables are equilibrated. Partially equilibrated states are described by Boltzmann-Gibbs distributions restricted to the set of configurations belonging to specific state (in this case the folded state).

In presence of strong friction effects we would expect the green and orange (and the corresponding work estimations) to differ. The total work along the cycle can be measured taking the difference between the force along the slow and fast part of the protocol. These quantities are shown in Fig S4C (green line shows the estimation based on the partial equilibrium force and orange line shows the direct measurement). In Fig S4D we show the estimated work obtained by numerical integration of the curves in Fig. S4C. for different values of the upper integration limit $\lambda$:

$$W = \int_0^\lambda (f_{RE} - f_{FW}) d\lambda \ . \qquad (S4.1)$$

The estimation shows that the two estimates are consistent within the experimental error, showing that friction effects are negligible. This is further illustrated in Fig S4E, that shows how for the condition $P_0 = P_1 = 1/2$, the Landauer limit is reached by increasing the pulling speed (Red data low speed, green data medium speed, blue data fast speed, dashed line Landauer limit).

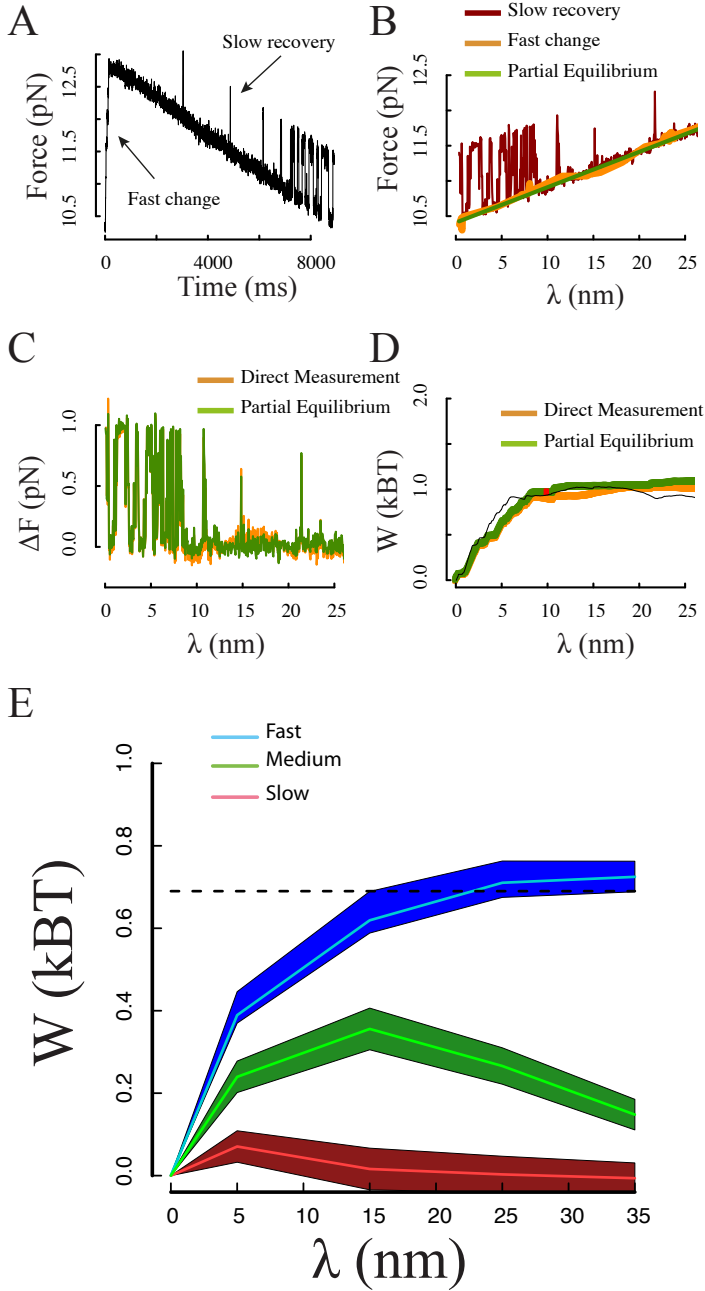

**FIG S4:** *A) Pulling protocol during work extraction showing the first fast change followed by the slow recovery. B) Pulling trajectory as a function of the control parameter λ. Orange line shows the fast change, red lines shows the slow recovery. Green line shows the estimated equilibrium force in the unfolded state. C) Force difference between the fast and slow part of the trajectory. The integral of this quantity gives the work extracted in the cycle. In orange we show the estimate based on the measured force and in green the estimate based on the partially equilibrated force in the unfolded state. D) Work estimates based on direct measurement (orange) and estimate of the partially equilibrated force (green). The two estimates are compatible within the experimental error. E) Average work extracted for different pulling speeds. High speed (blue data), medium speed (green data), low speed (red data).*

The extent of friction effects can also be estimated by introducing a simple model. In the absence of hopping the system can be effectively approximated as a series of two springs (see Fig. R6). When the trap-pipette distance is increased at constant speed $v$ the equation of motion for the bead position, $x$, reads:

$$\gamma \dot{x} = -k_M x - k_T(x - \lambda) + \eta \quad \quad (S4.2)$$

where η is a Gaussian white noise with correlation $<\eta_t \eta_s> = k_B T \gamma \delta(t-s)$. Moreover $k_M$ is the stiffness of the molecular tether and $k_T$ is the stiffness of the optical trap. Taking the average on both sides gives,

$$\gamma <\dot{x}> = -k_M <x> - k_T(<x> - \lambda). \quad \quad (S4.3)$$

We consider the case of constant pulling speed i.e. $\lambda = \lambda_0 + vt$, which can be solved as:

$$< x_t > = [R\lambda_0 + Rvt] + e^{-\omega t}(x_0 - R\lambda_0) + \frac{vk_T}{\omega(k_M + k_T)}(1 - e^{-\omega t}), \quad (S4.4)$$

where $\omega = (k_M + k_T)/\gamma$, $R = \frac{k_T}{(k_M + k_T)}$ and $x_0$ is the initial condition. The first term between brackets corresponds to the equilibrium solution, $x_{EQ} = [R\lambda_0 + Rvt]$. The position of the bead relates to the force measured in the optical trap as:

$$< f > = k_T(< x > - \lambda), \quad (S4.5)$$

and the total contribution of friction to the overall dissipation along the extracting cycle can be estimated by integrating the difference between $f$ above ad the equilibrium force $f_{EQ} = k_T(x_{EQ} - \lambda)$:

$$W_{DISS} = \int_0^{\Delta\lambda}(< f > - f_{EQ})\, d\lambda \approx \frac{k_T v R}{\omega}\Delta\lambda \quad (S4.6)$$

where transients have been neglected. In our system $\omega \sim 3\text{ ms}^{-1}, R = \frac{1}{20}, v = 2\frac{nm}{ms}, k_T = 0.06\frac{pn}{nm}, k_M = 1\text{ pn/nm}$ and $\gamma = 3\text{ pN ms/nm}$ so:

$$W_{DISS} = 0.04\frac{pn}{nm} \approx 0.01\, k_B T, \quad (S4.7),$$

well within our experimental error.

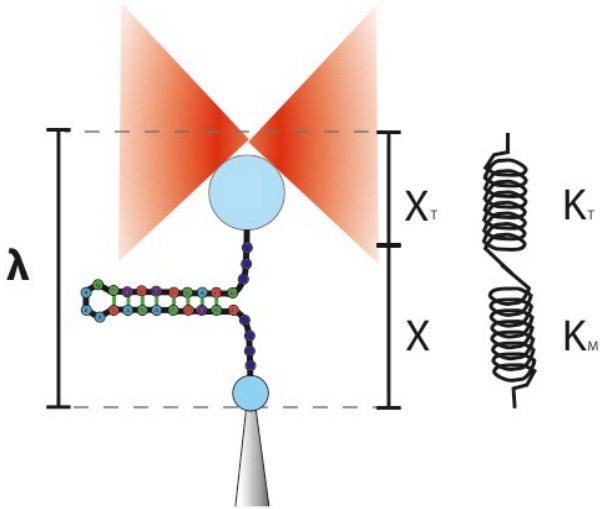

*FIGURE S5: Linear modelling of the experimental system. In the absence of folding/unfolding transitions the system can be modelled as a series of two springs. Here $k_M$ and $k_T$ are the stiffness of the molecular tether and of the optical trap respectively.*

**S5 : Key qualitative features of the CMD as a work-extraction machine**

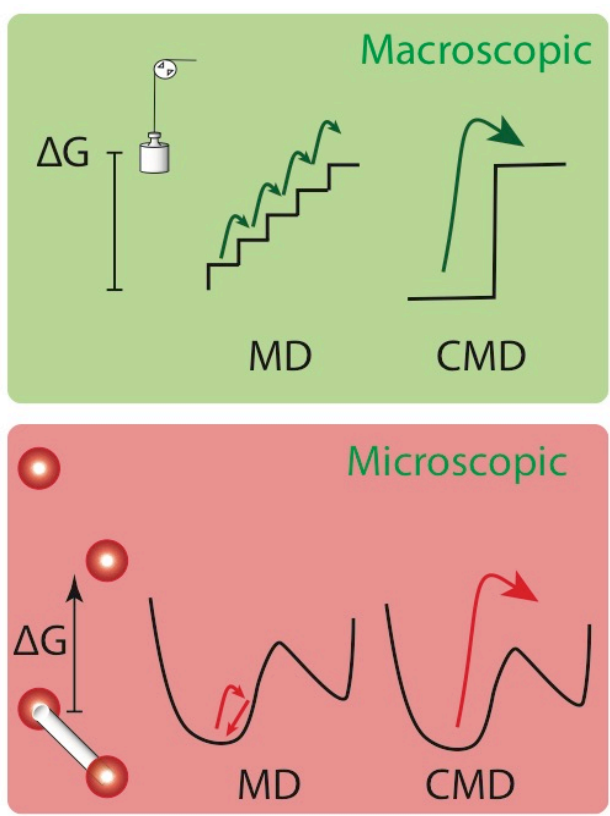

*Figure S6: Macroscale v.s. microscale. At the macroscale thermodynamic transformations can often be performed using multiple cycles each delivering a small amount of work. At the microscale, e.g. in enzymatic reactions, a fixed amount of work must be delivered in a single cyclem which is possible using a CMD but no a MD*

The remarkable aspects of the CMD are the following:

1) The CMD can extract arbitrary large amounts of work per cycle. In contrast the Szilard engine is limited to an average of $k_B T \log 2$ per cycle.

This is an essential point, especially at the nanoscale. In a macroscopic setting we can always imagine performing a given transformation (e.g. lifting a weight, Fig. S6 upper panel) either by delivering small amounts of work several times or by one single application of a large amount of work. At the microscopic scale the situation is quite different. For example, the enzymatic breaking of a single chemical bond in a single enzymatic cycle often requires the expenditure of large amount of energy, that might well exceed $k_B T \log 2$. As a consequence, if Maxwell-Demon-like devices are to be found in nature or used in technology, they must be able to deliver a specific amount of work per cycle. This can be achieved with information-to-energy conversion devices which store multiple bit sequences, rather than just one-bit devices such as the standard Szilard engine.

2) The efficiency of our scheme increases with the amount of work delivered per cycle. Notably efficiency does asymptotically reach 100% in the limit $P_0 \to 0$ or $1$, a regime dominated by rare events. This limit matches the efficiency of the standard Szilard engine while delivering more work per cycle. It is in this sense that we consider the CMD being able to exploit rare events better than the MD does.